**Title:** Entrepreneurship: some considerations

**Name:** Vítor João Pereira Domingues Martinho

**Affiliation:** Polytechnic Agricultural School of Viseu

**Address:**

Quinta da Alagoa - Estrada de Nelas

Ranhados

3500 - 606 VISEU

e-mail: vdmartinho@esav.ipv.pt

# Entrepreneurship: some considerations

**Abstract:**

In this work it is presented some considerations about entrepreneurship. Most of these questions are linked with Portuguese context. Portugal has some particularities, namely because the asymmetries between the littoral and the interior. This situation carried out some problems that complicate and prevent the appearance of new innovated business. In a situation of crisis like that we have today this context can become a really problem to solve some questions.

## 1. Introduction

The European Union as a whole urgently needs entrepreneurs to be more competitive (European Commission, 2010). It's a good sign the fact that most young people show interest in pursuing a career in entrepreneurial future. However, it remains concerned that the trends have not undergone major changes since 2004, particularly as regards the willingness to opt for self employment over employment for others.

The northern countries of the European Union are more interested in being entrepreneurs, unlike the southern countries, not to mention in the CEECs (Central Europe and Eastern Europe) which clearly show several obstacles to be entrepreneurs. The fact that the northern countries are more willing to take risks and be entrepreneurs have much to do with his capacity for organization and with his rigorous functioning of economies and societies. In a society lax in its organization, the entrepreneurs as good as they will always be difficulty in enforcing their business.

Only recently talking with a friend he told me that the northern countries of the European Union are truly an inspiration to us all and we would have much to learn if we spent there and occupy some of our time to analyze their styles life and how they overcame difficulties older. He told me that my friend actually many of us end up going through those countries, just to realize how they are organized and function, but then we have great difficulty in applying their reality in our day to day, because what they practice is part of the collective will and not of individual wills. Then we can also go there and bring what suits us and ignore what does not, trying to make a collage, to succeed, our way.

For these reasons, we considered relevant to present the twenty two approaches as set out below (Martinho, 2011).

## 2. The interconnection between economic sectors

Portugal in 2007, had about 1.2 million businesses, with 3.8 million employees and a turnover of around EUR 350 million (INE, 2010). The trade and services sectors are the largest number of firms and employees. Moreover, the industry has fewer firms, but has proportionately more people employed and a higher turnover. This is clear demonstration of the importance of industry to the development of any territory, or a region, country, or an association of countries such as the European Union. Moreover, there are several economic theories that link the industry as the sector's economic engine. Keynesians pointed to the industry as a sector with growing economies to scale, with capacity to produce tradable goods (which can be exportable) and internal dynamics that can positively influence other sectors of the economy. Therefore, the industry appears to be a good sector for economic development and other sectors appear to have also its importance.

In areas where development is more difficult, as is the case in rural areas, is the interconnection between the various sectors of the economy that may arise in new business. So today most of the European policies directed to rural areas come here to promote the development of an integrated manner, ie, to promote the emergence of businesses that are able to combine agriculture with the environment, and agriculture with tourism, or even agriculture with industry (not least the food industry). And in these territories is urgent new ideas appear, so the good preparation of action plans with public funding is essential.

Sometimes good ideas are close to us and in many cases just a few days we travel by train in Europe to see how our European neighbours do for many sectors of the economy.

### 3. The ability to keep companies created

In 2007, in Portugal, most of the companies were created in the services sector and the lowest in the industry (INE, 2010). This, beyond what has already been said on this subject, has much to do with the ease with which generate business services, many with small investments, unlike the challenge in implementing ideas in the area of industry that is a sector usually associated with more substantial investments.

But it is difficult to find a good idea and then implement it, is no less painful to keep it and make it back for periods more or less conducive to the planning of the business. In 2007, in Portugal, only 72.8% of firms set up is that managed to survive the first year, 53.8% survived until the end of 2nd year and only 47.1% that is able to reach the end of the third year. Here turns out to need to come good ideas, planning them very well and then go along with conviction. A positive attitude on these things can be the most crucial factor for ideas to remain and succeed.

If we analyze survival rates by sector notes that the industry is the sector with the highest survival rates in the first year and construction with the highest survival rates to 2/3 years. Services are the sector of the economy with lower rates of survival. This is again a reflection of what has been said above, ie the industry, unlike the services, typically require larger investments and so when it moves forward with an idea and creates a company it is usually well planned because what is at stake are more or less reasonable values that can not be applied with some lightness and so the business eventually implemented under the plan have better results.

## 4. High growth companies

The rates of business creation at the sectoral and spatial terms, can be a good indicator, but can not be analyzed separately. Should be complemented, particularly with the growth rates of business. Business growth can be measured both in terms of number of persons employed as paid through the turnover. The high growth companies are those that achieve an average annual growth over a period three years of existence, more than 20%.

If analyzed data for 2007, the growth of companies by number of employees paid the services sector is the one with highest number of firms with high growth both in number and percentage for the remaining (INE 2010). While it is analyzed by turnover growth the industry shows the largest number and construction companies the largest percentage. This fact is indicative of the characteristics and the relative importance of each sector.

It's much easier for companies related to services grow rapidly in terms of personnel, but it's harder to get it in terms of turnover. Moreover, industry is easier to grow in terms of turnover than in terms of number of employees. There is also the values listed firms from industry high growth represent a lower proportion compared to the other, compared to other sectors. Signal to have annual growth rates exceeding 20% over a period of three years is not easy in this business sector.

There is actually a set of interesting features in each sector of the economy that is worth knowing and which are of crucial importance at the time for decisions.

## 5. The regional problem

The evolution of the population in many different levels, is crucial developments in the economy, as has been said, and may generate a context of cause and effect at the level of regional disparities.

In Portugal, according to 2005 data, population growth has been guaranteed by international migratory movements that have not been sufficient to prevent the decline of people in the Alentejo region. The Açores were the only region that ensured the increase of people at the expense of a positive natural balance, ie, births exceeded deaths (INE, 2010). In 2005 the Norte had 35% of the population, the Centro 23%, Lisboa 26%, Alentejo 7.2% and other regions proportions less than 4%. The Lisboa region beside has a reasonable population proportion, almost to the level for example the Norte, has a population five times greater than this region.

This calls us to an issue that already is recurrent and that has to do with the saturation of coastal and desertification of the interior and the fact that the Lisboa increasingly attracting population and economic activity. At the bottom Lisboa gradually confirmed as a center of population and economic growth. This ought to be regarded by the authorities as a real phenomenon, and from there it was fitting that he or she decides to live with him or trying to alleviate. Because you can live with it and assume that most of the population and the activity goes to Lisboa and the remaining part of the territory is to the economic activity that can not be agglomerated (agriculture, some tourism, ...) and the inherent population . Or, think that this may not be good and has to take effective action, aggressive and appropriate to prevent such situations from happening.

Because in this context is the tendency for new business and new ideas appear mainly in Lisboa, since it has more people to think and generate new ideas and then more people and market to sell the good projects and good business.

## 6. The problem of education

According to INE (2010), data for 2004/2005, the Algarve was the only region which saw a national increase (2%) compared to the previous school year, in the number of students registred in obligatory education of nine years, divided into three consecutive cycles. Already in terms of retentions and abandonment in primary education (obligatory education) are what the Centro has the lowest percentage (10.1%) and Alentejo is the region with the highest weight in this phenomenon (13.9%). But if we look at retention rates and abandonment by cycles it appears that these will increase when we move in cycles, ie, the rates are higher in the third cycle (walking at around 19%) than in the first cycle ( plus or minus 5%).

At the secondary level are not very different context, namely, that Madeira is only a growth of the number of students registred (3.1%), with the Alentejo to present the greatest decline of students at this level of education (4 8%). The approval ratings are around 70% in all regions, with Norte to provide the best results.

In higher education about 55% of students were registred at this time in schools in Lisboa, Coimbra and Porto. On the other hand, the areas of health, science and business of engineering are preferred at the time of making decisions of choice of higher education for the future.

This is a brief portrait of what happens nationally, in educational terms, and quite reflective of the regional context that we have in these areas. If we think that some of the current economic theories, namely the theory of conditional convergence, point to the training of human resources as determinants of economic growth of societies and hence the creation of business and wealth, once again noted the difficulty some Portuguese regions.

### 7. The active population

In 2005 (INE, 2010) Portugal had 5.5 million working or unemployed but willing to work at the level of salaries, resulting in a participation rate of about 52.5%. Only the Norte and Centro had participation rates above average and the Açores had the lowest rate (45.4%). The female activity rate stood at 47.4%, having fallen in all regions below the national male. In young (15 to 24) this rate he walked around the 43.0% and always less than the total rate in all regions. Emphasis here is the Norte with rates above the national average.

By this time the workforce with a degree from the superior education corresponded to 13.2% of the workforce. Furthermore, over half the population had the third cycle of basic education, ie the current obligatory education. Lisboa was assumed as the region most qualified and less qualified Açores.

This year (2005) the unemployment rate in Portugal was already 7.6%. The Alentejo was the region that had unemployment rates higher, followed by Norte and Lisboa. Young people and women are the population segments most affected by unemployment.

Of the nearly 5 million who are inactive in Portugal about 57.1% were women, another part were students and about 12% were retired. The Alentejo is the region where the percentage of inactive pensioners is higher (46.5%).

These issues must be taken into account at the time to design strategies because they will be taken into account with certainty for economic agents who seek to make money. After this context also gives us an idea of business opportunities in each of the parts of the country. For example, who are working on business geriatrics at the Alentejo region will find good opportunities.

## 8. The distribution of economic activity

In 2003 (INE, 2010) the regions that contributed most to the GDP were Lisboa (38%), Norte (28%) and Centro (18.5%). At this time the Norte was the region which grew least (0.3%), unlike Madeira was the fastest growing region (5%) and the Algarve which grew by 3.5% over the previous year.

Lisboa, Norte and Centro concentrate a large part of economic activities, contributing in 2003 with 85% to national GDP. On the other hand, a finer level, for the NUTS III for example, notes that this time only the Grande Lisboa represents about one quarter of employment and about a third of the national productive activity. In addition, the Grande Lisboa had a GDP per capita above 70% the national average.

At the level of productivity Lisboa, Madeira and Algarve regions are above the national average. In sum is the coast and southern regions which have higher productivity levels.

Looking at the economic sectors are services that have higher values for productivity, derived from the very dynamics of financial intermediation and service businesses. The agricultural sector appears as the one with lower yields, resulting from much of the specific sector.

The agricultural sector (agriculture, forestry and animal) is characterized by having several features which make it different from other sectors but essential nonetheless. These particular issues related to price (there is a tendency to increase production faster than demand), markets (they are products of first need with elasticities price and income very specifically - people only consume what they can eat and are products of low unit prices), structural (many production units and small) and environmental (often by using excessive and indiscriminate).

Able to generate business in the midst of these regional and sectoral contexts, particularly in the less dynamic calls for the ability to find good ideas.

### 9. Business (a regional analysis)

During 2005 (INE, 2010) created in Portugal about 22 000 new enterprises, which indicates a decrease compared to 2004 which was evident in all regions of the country with the highest incidence in the autonomous region of Madeira and in the Alentejo region. The Centro was the one with less buoyancy in the formation of companies, counting this year with a ratio of only 5.2%.

Moreover, the number of companies dissolved walked around 16,000 that translate into an increase of 16% over 2004. The Norte, Centro and Alentejo regions that were a higher percentage of dissolution of companies, but were the autonomous regions of Açores and Madeira and Lisbon region that showed greater increases on previous years.

The density of business establishments showed, as would be expected, the highest values for metropolitan areas of Lisboa and Porto.

According to OECD criteria can identify five groups of activities related to manufacturing competitiveness in the light of that present, namely: natural resource-intensive industries, industries intensive in manpower, industries with high economies of scale, industries with great ability to differentiate their products and even industries intensive in R&D. Comparing the structure of regional GVA with the national average for each of these processing industries can find profiles of regional specialization. At this point in Portugal, the industries with the highest weight were the able to differentiate their products and the intensive in labour. The regions of Lisboa, Algarve and Centro were mainly characterized by manufacturing industries are able to differentiate their products. The Norte with the textile and footwear industries was characterized by intensive labour. Firms in the industry intensive in R&D were located mainly in Lisboa.

If we consider that firms are distinguished more by their ability to innovate, resulting in a lot of research, we conclude that there is still much to do in national terms, even more so when one knows the strategic importance of manufacturing industry.

## 10. The importance of international trade

The coverage rate of imports by exports in 2005 (INE, 2010) stood for Portugal in by 62%. In regional terms, the Centreo was the only region where the outflows exceeded those of entries (about 108%) and Lisboa was the one with the largest deficit. This positive development of trade, the Centro, it was also in the autonomous regions of Açores and Madeira, but especially the latter. The Norte, Centro and Lisboa were the regions that contributed most to international trade. The EU countries are the main interlocutors of the Portuguese regions in terms of international trade, while Spain is the country of choice for imports and exports.

Trade turn out to be determinants of the development (economic, business, etc.) from any drive space, especially when talking about exports to the country level. Moreover Keynesians point out that the demand by the export of tradable goods (goods produced by industry, mainly by manufacturing) is central to national development. Any country interested in having good levels of economic development must have a good bet in the manufacturing sector competitive and capable of produce good levels of export, either, minerals, textiles, footwear, automobile, chemical, food, high-tech. While Keynesians refer only to manufacturing industry with increasing returns to scale (mainly that uses high technology and little labor-intensive) is what is most interesting when thinking about having an aggressive and industry conditions to ensure high levels of exports. We also know that there are other factors that are driving the evolution of economies, but actually exports are still very crucial for countries if they can positively distinguish between them. In this sense Portugal still has a long way to go, especially when thinking about the need to innovate and be enterprising.

## 11. The role of agriculture and forestry

Taking into account such indicators as the number of farms, the utilized agricultural area, gross margins and annual work unit, and based on 2005 data from INE appears that in the last 16 years there has been a sustained trend of reduction in the number holdings (323 920) and the number of persons engaged (1.2 AWU per farm). It appears that, on the other hand, there was an increase in the average size of farms (11.4 ha on average) as well as the individual gross margins (8.3 thousand euros per farm).

As has been said the agrarian structure of the national land holdings characterized by a large number of farm and small. Therefore, it is not surprising that this time ¾ of the Portuguese holdings had less than five hectares. This structure is predominant in all agricultural regions, with the exception of Alentejo farms under five hectares account for only 45%. The Alentejo region is really a special case within the country in agricultural terms, as has the largest average area per holding (about 61 ha), with 9% of farm land and 45% of utilized agricultural area.

Regardless of the characteristics of production units, are agricultural or other sectors, opportunities, business, entrepreneurial and innovative and there are always possible, everything depends on the initiatives and the dynamics of economic agents involved.

There are many success stories in the national agricultural sector, referring to an example given in the website of the Association of Young Farmers of Portugal (http://www.ajap.pt/id.asp?id=s4sub14p) where they are referenced entrepreneurial companies from the production of rabbits to flowers, past the vineyards.

## 12. Venture Capital

Venture capital can be defined as a form of business investment, with the aim of financing companies, supporting their development and growth, with strong influence on management. Compared to other forms of financing, is the one that assumes the business success as the success of their own investment (CG International, 2010).

Usually the actors (all the people who organize themselves in order to enable the project) are the promoters of these investments (they offer entrepreneurs the

project), investors and management (team managing the project.)

In these operations it is possible to identify six stages: identification of the investor, analysis, research investment, and contract negotiation, implementation and operation of policies and incentives.

There are five types of structures through which invests in venture capital: SPAC (Special Purpose Acquisition Company), SGPS (holding company), IVC (Investor Venture Capital), SCR (Society for Risk Capital) and FCR Fund (venture capital).

The average citizen does not know or realize these socio-economic contexts and this is another factor that helps to lack of entrepreneurship in national terms. In sum we live in a society that are the financial rules that dictate the time of any and all business strategies. Know them and know how to play under those rules is crucial to the success of any business idea entrepreneur. Clearly there is still much to do in order to overcome these constraints. In not by chance that we live difficult days in Portugal and do not live a particularly good situation. It's funny that we have a common sense idea that if we pull together we have more power and economic theory confirms this.

### 13. Business Angels

The "Business Angels" manage the feat to raise money virtually fallen from the sky (CG International, 2010). The difficulties associated with the creation of businesses, mainly by young entrepreneurs, who often struggle with limited financial resources, are huge. The "Business Angels" can be a good help to overcome these initial obstacles.

This term is associated with the capitalists who, during the Great Depression, financed theatre in the "Broadway". Theatre that without this valuable assistance, those who were seen as business angels, could hardly have existed. It was thus for the historiography of the term "Business Angels" as a synonym for individual investors who bet on emerging businesses.

Structurally organized in Portugal since 1999, when it was created the first association, there are now about 350 "Business Angels" operating in Portugal throughout the various associations involved in this activity.

One of the associations is the Association of Portuguese Business Angels based in Lisbon and that counts as Associate Institutional BCC (Business Council of the

Centre). On the Internet site, this Association, which reads as a Business Angel is an investor who invests in emerging opportunities (like start-up or early stage). Participates in projects with smart money, ie, beyond the financial capacity dock, also contributing its experience and network of businesses. Business Angels have a number of characteristics in common, as are the investments that typically range between 25 000 and 500 000; like to exercise their capacity for mentoring project, seeking not only a high return on projects in which they invest, but also new challenges, preferably in your country or region.

Not always the lack of money can be an excuse not to carry out investment entrepreneurs.

### 14. The details make all the difference

Sometimes I find myself thinking how some businesses have grown so fast and so successful. Not knowing the business and its internal dynamics is often difficult to find a solid answer that is informative and objective of these facts.

Some days ago, however, I had the opportunity to meet with an entrepreneur known for whom I have great personal admiration. Only knew him by sight, had heard of him, had a meeting a few years ago with a team where he was present and had already given here to get an idea of the sources of his success stemmed and here much of its momentum and their way of being practical and pragmatic. But this week, as I said, I met with him and had the opportunity to visit the business facilities of which he is managing partner. On this visit he was impressed with how the facilities were designed and the fact that the smallest details and the smaller symbols were not left to chance.

Actually, judging by the success that this business has had the details make all the difference. Nowadays companies have to be distinguished from each other and it is here that we can do differently. Why are these things that most of us, in our day to day, bleach and think have no meaning, but maybe they have, can tell a lot and can affect the overall progress of a business.

So when you put the old question that is entrepreneurs born or are made to answer this question the more alive I have more difficulty in answering it. But I still believe that we can make good entrepreneurs, because the potential exists in each one of us, it is necessary to create the context and the means for people to develop their skills in the most accurate and profitable as possible.

### 15. Action Plans

We have unique natural resources continue to be underutilized. Here a few years ago I was in Spain to eat at a restaurant very famous in a small town, where one eats very well, but that by outward appearances, if I had been recommended, there would have hardly stopped. During lunch I spoke with the restaurant owner and he told me he was sorry we (Spanish and Portuguese) will not be all in a single Iberian country, because, for example, Portugal with the natural potential it has and with good organization Spain might be different today.

Actually we have unique natural potential and not just by the seaside part because these areas are of course historical reasons are more or less exploited. I refer especially to mountain areas and the interior that remain underutilized.

We lack good action plans that allow perspective the integrated and sustainable development of these areas. Very recently I participated in a debate on an area close to us and how best to leverage an endogenous feature very typical there. I enjoyed listening to a businessman linked to the wine sector to speak of the need of deprived areas to develop in an integrated and sustainable and that it should go through lesson plans that include, besides tourism and the promotion of endogenous resources, also bet on cultural programs that would attract people with some frequency.

At the bottom there are ideas, people with experience too, what is needed is to put all that rowing in the same direction and all focused on the same subject. Therefore, I have always been and remain an adherent of the platforms of understanding, whether sectoral or space and are more or less formal, allowing synergies together with the same objective. Imagine what it would create a platform of understanding for the agri-rural district of Viseu.

In times of crisis you have to innovate and invest in that has not been explored, because like others do and especially like the ones who are stronger than us, all we can is to fight with giants.

## 16. Agricultural Situation

According to data from INE (2010), forecasts a decrease in mass of the area sown of cereals autumn/winter, much because of adverse weather conditions, including the excessive rainfall. Moreover, the oil production shows signs of excellent prospects, both in terms of quantity and quality. Much the result of having bet truly aggressive policies aimed at increasing production, particularly of oil nationwide. In early 2010, sheep, cattle and pigs declines in the volume of slaughter at around 10%. Instead, we observed an increase in goats, which reached values of around 32%. The total weight of poultry and rabbits slaughtered for consumption was 22,863 tonnes, a rise of 5.2% over the same month in 2009. The production of cow's milk was collected from 149 000 tonnes, representing a decrease of 4.0% on amount collected in respect to that recorded in the corresponding month of 2009. The total volume of dairy products, however, increased slightly in January 2010 (+2.2%), again derived from a higher level of fresh products (milk and cream for consumption and acidified milk).

In early 2010, and compared with the previous month, the main changes in the index of producer prices were recorded for potatoes (+20.7%) in fresh vegetables (+7.2%), pigs (+ 4.9%), eggs (+3.5%) in olive oil (-8.6%) and in sheep and goats (-4.8%). It was noted, moreover, a slight positive variation of 0.7% in the price index of goods and services currently consumed in agriculture, while for the same period, the price index of capital goods does not record any variation.

With this context, at the level of production and prices, we say that coming interesting conditions to business in some areas, because producer prices are showing a trend of increase and prices of inputs are with the tendency to stagnation. Of course the price development over the years, changing production conditions and some structural issues may be decisive.

## 17. Tests of entrepreneurship

Easily we can find on the Internet sites that provide analysis tests of the entrepreneurship capacity of each one of us. One of these sites is, for example, the test IAPMEI where consider those to be key characteristics of an entrepreneur, namely, independence, self-discipline, creativity, motivation, ability to risk and confidence.

Independence is essential for anyone who wants to develop an entrepreneurial initiative. Hardly anyone who systematically relies on third parties to make a decision will be able to take risks and come up with an innovative idea.

Being able to self-discipline is also a key feature for an entrepreneur, especially in terms of time management, given the enormity of requests that we face every day.

Likewise, creativity is crucial because only with it we can have innovative ideas and different from existing ones.

Be persistent, motivated and confident is of paramount importance, because first we must be able to sell an idea to ourselves, because if not we can not sell to anyone else.

The information available on the website of IAPMEI states that the documents on Entrepreneurship in Europe stress that the most important aspect for all those involved in the processes of promotion of entrepreneurship is the assessment of entrepreneurship capacity of the people. It also states that there are not people with entrepreneurial skills innate and that much of what is necessary for the success of a business can be obtained through an ongoing learning process. However, there are a number of inherent aspects that can make this process more feasible. Although, as I have said many times, opinions about whether people are born, or become, entrepreneurs are not unanimous.

### 18. The entrepreneurial activities and business efficiency

In the Journal of Entrepreneurship (2010) we can find several interesting and topical scientific articles on issues of entrepreneurship. One of these articles has to do with relations between the entrepreneurial activities of national and regional development and business efficiency. This work has been based on the context of Croatia and I find interesting to examine it and comment it here, because it shows clearly that not always what seem interconnected, it is in the reality.

In concrete terms, the study examines the dynamics, structure and connections between business, economic development and business efficiency. The authors conclude from this study that, although we think in terms of common sense that the relationship between these variables has direct effects, linear and positive, the empirical evidence shows that the impacts are really significant, but are complex and few direct.

Evidence of entrepreneurial activity in Croatia show that the initial phases of the development of entrepreneurial activities are very dynamic, but also very volatile. In addition, there are important regional differences in entrepreneurial activity in business performance and economic development among the six regions of Croatia. The correlation between entrepreneurial activity, business performance and economic development are important, but depend on whether the business is based on chance or necessity. In other words, this study confirms the theoretical assumption about the complex and multifaceted connections between different types of entrepreneurial activity and economic development.

In short, entrepreneurship is not established by the law and by utopias, must be take into account the national contexts of each country, region, etc.

### 19. Innovation in small and medium enterprises

Innovation in small and medium enterprises is a growing concern of its leaders, because only in this way they can reach the ultimate goal of any manager who is having the monopoly of its production. Thus, they can eliminate competition and can define rules to the market where they produce. A phase of crisis like the one we currently have these are aspects that can make big differences. The website of the Journal of Entrepreneurship has an interesting scientific article on the subject of innovation in small and medium enterprises worth to pull here, because often there are structural aspects that make the development of certain contexts not linear.

Specifically, the objectives of the study are: to assess the impact of individual characteristics, available resources, organizational culture, structure and market dynamism in the innovation of small and medium enterprises, developing knowledge for innovation in enterprises. The research methodology follows the method of approach to the case of Yin (1994). In these case studies involved two companies in Information and Communication Technology (ICT), winners of several awards for innovation in the UK. The evidence from these studies shows that the intensity of innovation is dependent on the availability and adequacy of human and financial resources. In addition, an organizational culture to support the development of new products and an infrastructure conducive to innovation will positively influence the outcome of innovation. The results also showed that innovation at company level is influenced internally by the experience of the management and externally by the dynamism of the market.

So I would say that is entrepreneur not who want, but who can.

## 20. Paradigm shift

The emergence of new ideas of business is much related with the shift of paradigm, in other words with the change of what we are accustomed to seeing. I refer, for example, two potential business ideas that go through by the change of the usual way how we look at products that we every day use, like the solid oil to spread and fruit bars packed as cereal bars or chocolate.

The project "solid oil to spread," presented to the Innovation Agency by LET-ISA, together with Consulai, a consulting firm in the food and agri-food, and Cooking.Lab, an expert in molecular gastronomy, winner of the Innovation Award BES for the year 2008 in the Agro-Industrial category. It is hoped that this new product will be available on the market very soon. This is one of the most prestigious awards for innovation in Portugal, and this was the first time that the BES in the competition included the agro-industrial category.

The project bars of fruit packed is a potential business idea that I saw being presented in the program SCI News, "Egg of Columbus" a few days ago. Bars of fruit packed can be the next healthier snack to arrive to the market. Is a Portuguese project but will be sold first in countries with high consumption. Industrial innovation is a way to maintain the taste and smell.

These are, in short, two ideas that seem fabulous, with the potential to generate much business and have always been under our eyes, the difference was in giving an alternative use. It is with these ideas that we need to change the Portuguese economic issue and at least generate employment.

## 21. Social initiatives

I see around in the Internet to research information about entrepreneurship and I found, in a lot of information, reference to various initiatives under the social

entrepreneurship, and I gave attention to an initiative called "New Settlers."

According to information provided on the page of this initiative, "New Settlers" is a development project that enhances opportunities for the reduction of regional disparities. In the background to help low-density regions competing with overpopulated urban areas. Want to target those who are awake to go from an everyday urban, for a quieter life in rural areas, with an agenda that facilitates the integration of "new settlers" in the new reality. This concept aims to induce a new dynamic in areas of low density. An initiative in the line of an office for attracting investment in this case, attracting new residents entrepreneurs with features capable of inducing significant synergies within the territories and becoming this zones more competitive.

Yet in this Web page I found various expressions that I think is important to reproduce here: lose the fear of failure; entrepreneurship is not employment, but lifestyle; entrepreneurs do not just want to impact the world, but also receives money; take a chance on their dreams, even if it means that you might fail; more than 1.5 million people earning a living selling things on eBay; governments, entrepreneurs and employees should not fear risks, because they make us winners auspicious; the base of the pyramid may be the source of new business cycles, if we embrace the risk; people are the best kind of capital; if there is no market, no deal; fail is simply the opportunity to begin again, this time more intelligently; people think that entrepreneurs are those who can not find jobs, but I had one; even large groups and traditional had a beginning entrepreneur; and what, if we fail? we learned something, maybe as much as in college; crash is the most intensive form of doing market research.

**22. Innovative uses of food waste**

Searching in web for examples of entrepreneurship in food subjects and I found one blog about entrepreneurs, the description of one of the most recent initiatives created at the University of Aveiro, "IFoodTech - Technological Ingredients for Food". According to information available on this blog about it, this initiative aims to analyze the potential of some agro-food sub-products such as derived from the coffee and grape, to produce new products with added value.

Made with non-polluting technology, refers to the blog, the Investigation Unit of Organic Chemistry of Natural Products and Agro-food, of the University of Aveiro, the new components are extracted with low value, may have uses in products for the benign health and welfare.

With the scientific knowledge of the group of researchers, it may be possible to develop techniques for extracting food compounds and give them new applications, such as making healthier food, as, according to the blog, says the researcher responsible. Our goal, says the researcher, will be identified rich sources of natural compounds that can easily be included in various types of foods, extract them and give them new applications. Different properties, that chemists and biochemists, can exploit to face with future applications.

This is another good example of how research and science have been made to serve the economy, wealth creation and promoting employment. Much more could be done in various parts of the country if we can create more synergies between the different institutions.

### 23. The soul is the secret of business

It is popular adage that the secret is the soul of business, but I, to paraphrase a well known and successful entrepreneur of our area, I would say that is the soul that is the secret of the business. When we say that we must be different from others, we soon find innovative ideas and very fantastic, what is also important, but sometimes is enough change our attitude and our soul for just here arrange a difference.

We Portuguese have a characteristic that distinguishes us from most other people that is the spirit of sacrifice. If we can take this and other features for sure we will make a difference. I would say that before we start thinking about making great adventures we will be able to leverage these characteristics that are much ours.

Other people with labour cheaper could practice lower prices and may copy technologies, but will be unable to copy our characteristics, at the bottom of our soul. In our daily lives we must convince ourselves that we are able to achieve what we want and if so, aware that at the middle will always appear immense difficulties, our objectives are achieved.

It is also true that almost nothing prepares us to be tenacious, determined, stubborn and determined, nor the education system from an early age, nor society. Anyway we must change what is wrong, print that soul needed and stop wasting time with wailing that the fault is always of the other, when each of us may at any time to contribute to the improvement of the environment.

## 24. Conclusions

Speak and talk about entrepreneurship is not easy because is a emergent issue, but it is left here some considerations about this issue. It is stressed namely the Portuguese situation. Taking into account the economy theory, principally convergence theory, new economic geography and Keynesian theory the interior of Portugal will have very hard difficulties to find new business. There are also some economic sectors with the same problems.